\documentclass[aps,prb,twocolumn,groupedaddress,showpacs,superscriptaddress,amssymb,amsmath]{revtex4-2}
\usepackage{natbib}
\usepackage[utf8]{inputenc}
\usepackage{graphicx}
\usepackage{tabularx}
\usepackage{xcolor}
\usepackage{amsmath}

\usepackage{comment}
\usepackage{dcolumn}
\usepackage{hyperref}
\hypersetup{
	colorlinks=true,
	linkcolor=blue,
	citecolor=blue,
    urlcolor=blue
}

\usepackage{bm}
\usepackage{epsf}
\usepackage{braket}
\usepackage{tensor}
\usepackage{soul}
\usepackage{mathrsfs}
\usepackage{mathtools}
\usepackage{multirow}

\usepackage{amsfonts}

\newcommand{\be}{\begin{equation}}
	\newcommand{\ee}{\end{equation}}
\newcommand{\bea}{\begin{eqnarray}}
	\newcommand{\eea}{\end{eqnarray}}

\makeatletter
\newsavebox{\@brx}
\newcommand{\llangle}[1][]{\savebox{\@brx}{\(\m@th{#1\langle}\)}%
	\mathopen{\copy\@brx\kern-0.5\wd\@brx\usebox{\@brx}}}
\newcommand{\rrangle}[1][]{\savebox{\@brx}{\(\m@th{#1\rangle}\)}%
	\mathclose{\copy\@brx\kern-0.5\wd\@brx\usebox{\@brx}}}
\makeatother

\begin{document}



\title{Principal component analysis of wavefunction snapshots in non-equilibrium dynamics}

\author{Dharmesh Yadav}
\email{dharmesh.yadav@students.iiserpune.ac.in}
\affiliation{Department of Physics,
		Indian Institute of Science Education and Research, Pune 411008, India}

\author{Devendra Singh Bhakuni}
\email{dbhakuni@ictp.it}
\affiliation{The Abdus Salam International Centre for Theoretical Physics, Strada Costiera 11, 34151 Trieste, Italy}
        
\author{Bijay Kumar Agarwalla}
\email{bijay@iiserpune.ac.in}
\affiliation{Department of Physics,
		Indian Institute of Science Education and Research, Pune 411008, India}
\begin{abstract}

We study non-equilibrium quantum dynamics by performing principal component analysis on the data sets of wavefunction snapshots.  We show that a specific transformation of the data sets maximizes the information content in the largest principal component and further enables its connection to certain observables.  This connection enables us to explain the dynamical features revealed by such a dimensionality-reduction scheme. We demonstrate this using quantum dynamics of the Heisenberg spin chain, starting from different initial states, and further extend the approach to extract higher-order correlations. Our framework should also be applicable to other unsupervised machine-learning methods based on dimensionality-reduction schemes and is highly relevant to experiments with quantum simulators, including those in higher dimensions. 

\end{abstract}
  \maketitle  

{\it Introduction.--} 
Quantum simulations of many-body interacting systems across diverse experimental platforms have emerged as a powerful paradigm for exploring complex quantum phenomena, ranging from quantum phase transitions \cite{A_quantum_Newtons_cradle,AHAMI2024129770,Quantum_simulation_of_quantum_phase_transitions, Sachdev_2011}, non-equilibrium dynamics and associated universality \cite{Non_eqb_dynamics_1/3_universality,Non_eqb_dynamics_Absence_of_Diffusion_in_an_Interacting_System_of_Spinless_Fermions_on_a_One_Dimensional_Disordered_Lattice,Non_eqb_dynamics_Anomalous_Diffusion_and_Griffiths_Effects_Near_the_Many-Body_Localization_Transition,Non_eqb_dynamics_Anomalous_dynamic_and_equilibration_in_the_classical_Heisenberg_chain,Non_eqb_dynamics_Comment_on_Anomalous_spin_diffusion_in_classical_Heisenberg_magnets,Non_eqb_dynamics_Detection_of_Kardar_Parisi_Zhang_hydrodynamics_in_a_quantum_Heisenberg_spin-1/2_chain,Non_eqb_dynamics_Dynamic_Scaling_of_Growing_Interfaces,Non_eqb_dynamics_Dynamic_scaling_relation_in_quantum_many-body_systems,Non_eqb_dynamics_Emergent_Hydrodynamics_in_Integrable_Quantum_Systems_Out_of_Equilibrium,Non_eqb_dynamics_Entanglement_in_a_fermion_chain_under_continuous_monitoring,Non_eqb_dynamics_Evidence_of_Kardar-Parisi-Zhang_scaling_on_a_digital_quantum_simulator,Non_eqb_dynamics_Family-Vicsek_Scaling_of_Roughness_Growth_in_a_Strongly_Interacting_Bose_Gas,Non_eqb_dynamics_KPZ_dynamics_in_Integrable_Quantum_Systems,Non_eqb_dynamics_KPZ_universality_one-dimensional_polariton_condensate,Non_eqb_dynamics_Long-lived_solitons_and_their_signatures_in_the_classical_Heisenberg_chain,Non_eqb_dynamics_Nonlinear_Fluctuating_Hydrodynamics_for_Anharmonic_Chains,Non_eqb_dynamics_Nonlinear_Fluctuating_Hydrodynamics_for_the_Classical_XXZ_Spin_Chain,Non_eqb_dynamics_Quantum_gas_microscopy_of_Kardar-Parisi-Zhang_superdiffusion,Non_eqb_dynamics_Scale-invariant_critical_dynamics_at_eigenstate_transitions,Non_eqb_dynamics_Scale-Invariant_Survival_Probability_at_Eigenstate_Transitions,Non_eqb_dynamics_Spin_transport_in_a_tunable_Heisenberg_model_realized_with_ultracold_atoms,Non_eqb_dynamics_Time-dependent_behavior_of_classical_spin_chains_at_infinite_temperature,Non_eqb_dynamics_Universal_spin_dynamics_in_infinite-temperature_one-dimensional_quantum_magnets,RevModPhys.83.863,Gogolin_2016,D_Alessio03052016,fermion_bipartite_fluctuations}, quantum transport in condensed matter and statistical physics to the simulation of lattice gauge theories from high-energy physics \cite{quantum_simulation_lattice_gauge_theories,Quantum_simul_Schwinger_model,Quantum_simulations_of_gauge_theories, Floquet_approach_Quantum_simulations_of_gauge_theories}. Crucially, such platforms provide access not only to local observables but also to full snapshots of the many-body wavefunction \cite{KPZ_higher_order_superdiffusion, Snapshot_Local_Readout_and_Control_of_Current_and_Kinetic,Snapshots_emergent_hydrodynamics_long_range_magnet,Snapshots_51-atom_quantum_simulator,Snapshots_hydrodynamics_chaotic_QS,Snapshots_Q_gas_microscopy_single_atom,Snapshots_Many-body_physics_with_controlled_Rydberg_atoms}. These capabilities enable the investigation of full counting statistics, nonlocal correlations \cite{Snapshots_hydrodynamics_chaotic_QS,joshi2025measuring,Non_eqb_dynamics_Quantum_gas_microscopy_of_Kardar-Parisi-Zhang_superdiffusion,rosenberg2024dynamics}, and the properties of measurement outcomes (“bit strings”), providing new insights into concepts such as Hilbert-space ergodicity and deep thermalization~\cite{Cotler2023emergent,Ho2022exact,Mark2024maximum,shaw2025experimental,Ippoliti2023Dynamical}. 

Access to large datasets of many-body wavefunctions further brings data science and machine learning approaches into play, enabling the extraction of information from full datasets in a largely agnostic manner, without relying on detailed prior knowledge of the underlying system. Although both supervised and unsupervised machine-learning methods have been successful in contexts such as predicting equilibrium phase transitions \cite{Discovering_phase_transitions_with_unsupervised_learning, Discovering_phases_phase_transitions_and_crossovers_through_unsupervised_machine_learning, Unsupervised_learning_of_phase_transitions_From_principal_component_analysis_to_variational_autoencoders, Unsupervised_Machine_Learning_of_Quantum_Phase_Transitions_Using_Diffusion_Maps,andreoni2025networktheory,ziv2025unsupervisedmachinelearningexperimental} and uncovering universal aspects of nonequilibrium dynamics \cite{Data-driven_discovery_statistically_relevant_information_in_quantum_simulators, PCA_absorbing_phase_transitions, Machine_learning_pair_contact_process_with_diffusion, machine_learning_based_compression_quantum, Learning_nonequilibrium_statistical_mechanics, Non_eqb_through_ANN, Supervised_learning_directed_percolation}, it remains unclear exactly where these methods are effective and what their fundamental limitations are. In particular, for out-of-equilibrium contexts, an unsupervised dimensional reduction scheme, based on Principal Component Analysis (PCA), captures the correct dynamical exponent corresponding to quantum transport when starting with a domain wall initial state \cite{Devendra_PCA}; however, its applicability for other generic, experimentally feasible initial states is not yet clear. Moreover, for any generic initial state, it is not clear
which observable mimics the dynamics of the largest principal component of the PCA analysis, and whether such a framework provides insights into non-local correlations.

In this work, we provide answers to these questions. Generally, when performing PCA on wavefunction snapshots starting from any initial state, the information can be spread across all PCA components, depending on the choice of initial state. Here, we show that by working properly with a transformed matrix, the information in the largest principal component can be maximized. Moreover, this transformation immediately provides information on the observable whose dynamics can be well approximated by considering only the largest principal component. We further extend our analysis to the case in which the largest principal component approximates the dynamics of higher-order correlations. We demonstrate our results using a prototypical example of the one-dimensional quantum Heisenberg spin chain, with dynamics starting from different initial states. 

\textit{General workflow and construction scheme.--} The analysis in this study is performed in the following steps: (1) dynamical evolution and measurement, and (2) construction of the snapshot matrix and performing PCA to extract useful physical quantities. 

\textit{(1) Dynamical evolution and measurement.--} The many-body quantum state is evolved starting with a desired initial state and Hamiltonian $H$. Here, we primarily focus on interacting spin systems, but the procedure can be suitably adapted for other types of systems as well. As the system evolves, dynamics are probed by performing single-site projective measurements along the $z$ axis on all $L$ sites. The measurement procedure gives a binary string \textbf{n}=$(n_1,..., n_L)$, $n_i \in \{0,1\}$ for $i=1,...,L$ called as \textit{wave function snapshots} at a particular time $t$, where the value $0$ is for spin down and $1$ for spin up. These snapshots are arranged as row vectors in a matrix $\mathbf{X}(t)=\{\textbf{n}^{(j)}(t)\}$, where $j=1,...,N_r$, are the different realizations \cite{Devendra_PCA}.

\textit{(2) Snapshot matrix and PCA.--} Next, we extract the information from the snapshot matrix using a dimensional reduction scheme based on PCA on the measured rectangular data matrix $\mathbf{{X}}$. The eigenvalue decomposition of a symmetric $L \times L$ matrix $\Sigma = \frac{\mathbf{X}^T \mathbf{X}}{N_r}$ is performed to find the PCA eigenvalues $\{\lambda_1,...,\lambda_L\}$. These PCA eigenvalues are arranged in increasing order $\lambda_1 \geq \lambda_2 \geq \cdot \cdot \cdot \geq \lambda_R>0$, where $R$ is the rank of the matrix $\mathbf{{X}}$. These PCA eigenvalues are also related to the singular values $s_k$ of the matrix $\mathbf{{X}}$ as $\lambda_k=s_k^2/N_r$. 

In what follows, we first perform PCA on the data matrix $\mathbf{X}$, which we call the bare dataset. We will show that, for certain initial states, the largest principal component, $\lambda_1$, already provides information about the underlying dynamics of some observables. However, there are cases where it fails to do so, and we provide an approach to maximize the weight on $\lambda_1$ and to further connect the dynamics of the largest component with certain observable dynamics.

We focus on the interacting XXZ spin 1/2 chain whose Hamiltonian is given as
\begin{equation}
    H = \sum_{i=1}^{L-1} J \big(S_i^xS_{i+1}^x+S_i^yS_{i+1}^y\big) + \Delta \sum_{i=1}^{L-1} S_i^z S_{i+1}^z.
    \label{XXZ-spin}
\end{equation}
Here, $S_i^k, k=x,y,z$ are spin-1/2 operators at site $i$, $J=1$ is the interaction strength in the $\mathrm{XY}$  plane, and $\Delta$ is the anisotropy parameter. The XXZ Hamiltonian conserves total magnetization along the $z$-axis, i.e. $[M_z, H]=0$, with $M_z= \sum_{i=1}^{L} S_i^z$. To analyze the behavior of the PCA eigenvalues $\lambda_i$ of the snapshot matrix, we study the quantum dynamics starting from three different initial states: the domain wall (DW), the N\'eel state, and an XZ-type multi-periodic domain wall (MPDW) state \cite{universalscalinghigherordercumulants, supp}.

\begin{figure}
    \centering
    \includegraphics[width=1\linewidth]{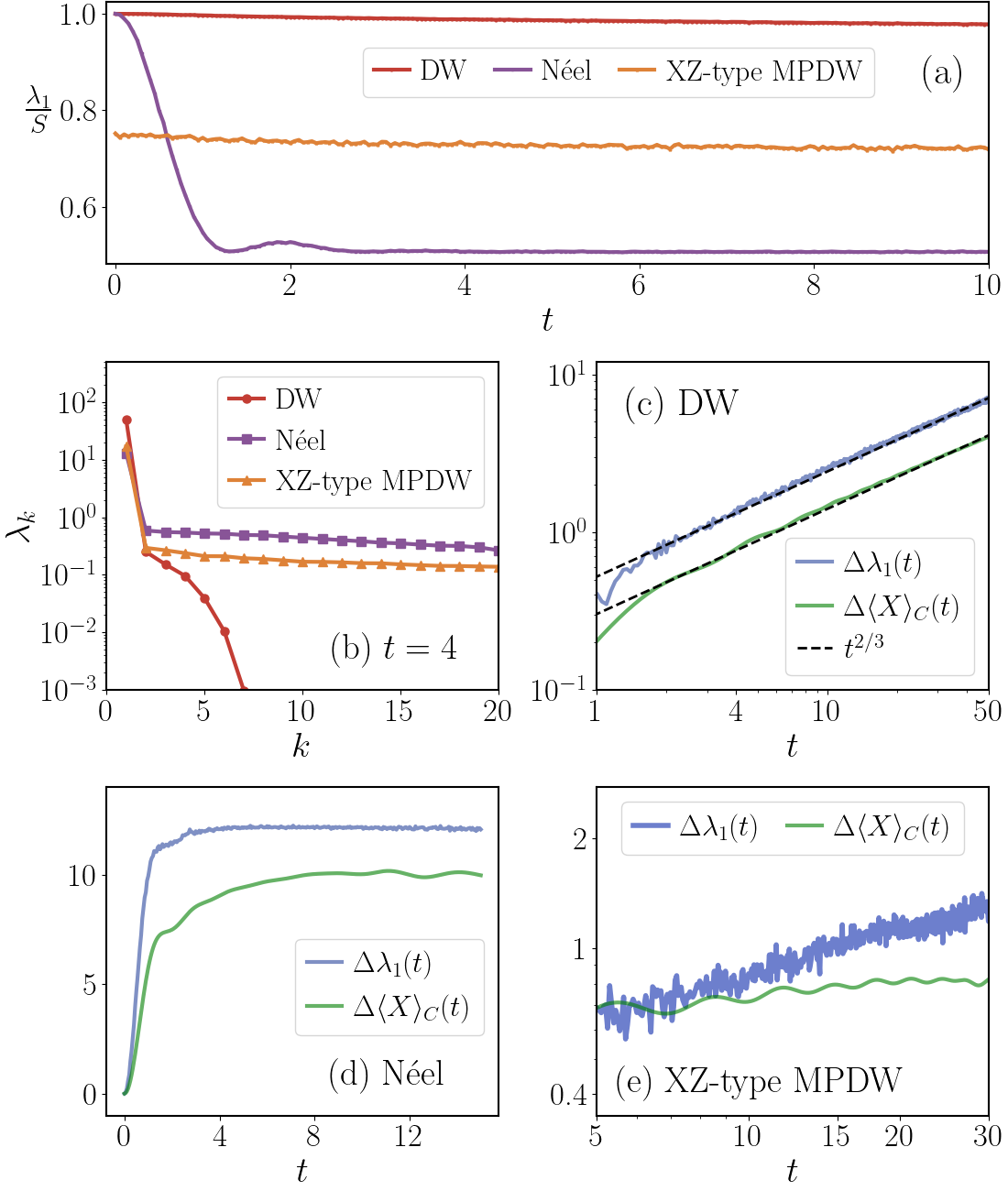}
    \caption{(a) Time evolution of the scaled eigenvalue $\lambda_1$ (scaled by $S=\sum_{i} \lambda_i$) for three different initial states:  (i) domain wall (DW) state which have two halves of the lattice with opposite spin polarization, (ii) N\'eel state, characterized by alternating down and up spins, and (iii) XZ-type multi-periodic domain wall (MPDW) state, which has a helical mixed-basis form composed of $x$ and $z$ bases with wavelength $\lambda_{XZ}$ \cite{supp}. (b) Decay of PCA eigenvalues $\lambda_k$ versus $k$ at a fixed time $t=4$ for the same three initial states. 
    (c)-(e) Dynamics of $\Delta \langle{X}\rangle_C(t) = |\langle{X}\rangle_C(t) - \langle{X}\rangle_C(0)|$, defined in Eq.~\eqref{x-eq}, and the principal component $\Delta \lambda_1(t)= |\lambda_1(t)-\lambda_1(0)|$ for the DW, Néel, and the XZ-type MPDW states, respectively.
    For the DW case, $\Delta \lambda_1(t)$ approximates the dynamics of $\Delta \langle{X}\rangle_C(t)$ quite accurately, whereas for the other cases, they carry distinct dynamical features. 
    For the numerics, we consider $\Delta=1$, $L=100$, $\lambda_{XZ}=100$, and $N_r=1000$ for DW and N\'eel, whereas $N_r=4000$ for XZ-type MPDW state. In plots (c)-(e) the quantity $\Delta \langle{X}\rangle_C(t)$ is multiplied by a constant factor $d$ for plotting it on the same scale as $\Delta \lambda_1(t)$. The value of $d$ is $1$ for (c), $40$ for (d), and $6.5$ for (e). Note that the mismatch between $\Delta\langle{X}\rangle_C(t)$ and $\Delta \lambda_1(t)$ in (d) and (e) is due to the contribution arising from higher eigenvalues of the snapshot matrix.}
    \label{fig:Bare_PCA}
\end{figure}

In Fig.~\ref{fig:Bare_PCA}(a), we plot the dynamics of scaled $\lambda_1$ (scaled by $S=\sum_{i=1}^{R}\lambda_i$) for the sub-system data-matrix $\mathbf{X}$ of size $N_r \times \frac{L}{2}$ (bare datasets) starting from three different initial states, as mentioned before.  As shown, for the DW state, $\lambda_1$ has the dominant contribution among all the initial states. In particular, for the N\'eel state, the quantity $\lambda_1/S$ quickly saturates to $1/2$ \cite{supp}, indicating faster spreading of information in the data set. In Fig.~\ref{fig:Bare_PCA}(b), we plot the spectrum of the snapshot matrix $X$ at a given time. The plot clearly shows that, for the DW state, $\lambda_1$ carries a significant fraction of the information compared to the remaining eigenvalues. In contrast, for both the XZ-type MPDW and the N\'eel state, the contributions from all the eigenvalues, other than $\lambda_1$, are of a similar order.

Moreover, different dynamical features in $\lambda_1(t)$ can be observed for different choices of the initial states: For the DW state, the dynamics of $\Delta\lambda_1(t)=|\lambda_1(t)-\lambda_1(0)|$ follows a power law growth, as shown in Fig.~\ref{fig:Bare_PCA}(c), while for the Néel state, it saturates at early times, as shown in Fig.~\ref{fig:Bare_PCA}(d). To connect the dynamics of $\Delta \lambda_1(t)$ with some observable, we compute the simplest observable that can be constructed from the data matrix, which is the classical average of the snapshot matrix of size $N_r \times \frac{L}{2}$, defined as 
\begin{equation}
\label{x-eq}
  \langle X\rangle_C =\displaystyle \frac{2}{N_r L}\sum_{i=1}^{L/2} \sum_{j=1}^{N_r} n_{ji}.  
\end{equation} 
For the considered classical encoding, the observable in Eq.~\eqref{x-eq} corresponds to the average magnetization (particle number) in a subsystem of size $L/2$. As shown in Fig.~\ref{fig:Bare_PCA}(c), for the DW state, $\Delta \langle{X}\rangle_C(t) = |\langle{X}\rangle_C(t) - \langle{X}\rangle_C(0)|$
captures dynamics similar to that of $\Delta \lambda_1(t)$, thereby producing super-diffusive scaling  $\Delta \lambda_1(t)\propto t^{1/z}$ with dynamical exponent  $z=3/2$ upto the time scales considered.  We note that for the DW initial state, diffusion is expected at long-times \cite{PhysRevB.96.195151}. This suggests that for the DW state, $\lambda_1(t)$ approximates the dynamics of the average magnetization in the subsystem. In contrast, for the other two initial states, as shown in Fig.~\ref{fig:Bare_PCA}(d)-(e), $\Delta \lambda_1(t)$ and $ \Delta \langle  X\rangle_C(t)$ have distinct dynamical features, and hence, for these cases, $\lambda_1(t)$ alone does not approximate the dynamics of $\langle X \rangle_C(t)$, indicating that a simple dimensional reduction is not always possible. In what follows, we propose a new modification to the snapshot matrix $\mathbf{X}$ that not only maximizes the weight on $\lambda_1$, but also allows us to express it in terms of the expectation value of some observables, thereby providing information about which dynamical features can be extracted by the largest principal component $\lambda_1$, or in general, by the PCA analysis.

\textit{Modified construction.--} We first note that the original snapshot matrix $\mathbf{X}$ contains variables measured in the $z$ basis. Hence, one can construct an arbitrary Hermitian operator ${O}$ by a linear combination of local $S_{i}^{z}$ operators as
\begin{equation}
    O = \sum_{i=1}^M a_i S_i^z,
\label{adjusted_magnetization_defn}
\end{equation}
where each site $i$ being weighted by a real scalar factor $a_i$ which can take two possible values $\pm 1$, and $M \leq L$ is the sub-system size. For $a_i=+1$, $O$ corresponds to magnetization operator, whereas $a_i = \mathrm{sgn}[\langle{S_i^z(t=0)}\rangle]$ corresponds to spin polarization, and $a_i = (-1)^i$ corresponds to staggered magnetization, within the domain of size $M$.

Depending on the choices of $a_i$, different transformations of the snapshot matrix $\mathbf{X}$ can be constructed as follows: Given the original binary string \textbf{n}=$(n_1,..., n_L)$,  $n_i \in \{0,1\}$ for $i=1,..., L$, we construct the new snapshot matrix $\mathbf{\overline{X}}$ consisting of $\mathbf{\overline{X}} = \{\overline{n}^{(j)}(t)$\}, $j=1, \cdots N_r$, 
\begin{equation}
   \overline{n}_i = n_i \oplus \overline{a}_i, \quad i=1, 2, \cdots M
\label{Transformed_snap_new_observable}
\end{equation}
where $\bar{a}_i$ is the computational space variable which takes the value $0$ when $a_i=+1$ and the value $1$ when $a_i=-1$. As a result, this transformation correctly maps the spin representation in the computational space. Here, we consider the subsystem of size $M$. Performing PCA on this modified snapshot matrix $\mathbf{\overline{X}}$ then allows relating the corresponding $\bar{\lambda}_1$ to the expectation value of the observable $O$ in Eq.~\eqref{adjusted_magnetization_defn}. This can be shown by using the fundamental theorem of singular value decomposition \cite{Principal_component_analysis_review}, which  relates the entries of a general matrix $\textbf{A}$ to its singular values $s_k$ as, 
\begin{equation}
||\textbf{A}||_F^2=\sum_{ij}\textbf{A}^2_{ij}=\sum_{k=1}^Rs_k^2,
\label{SVD}
\end{equation}
where the subscript $||.||_F$ refers to the Frobenius norm of the matrix $A$ and $R$ is the rank. As for $\mathbf{\overline{X}}$, the entries are only $0$ and $1$, the Frobenius norm of  $\mathbf{\overline{X}}$ can then be expressed as
\begin{equation}
\sum_{i=1}^{M} \sum_{j=1}^{N_r} \mathbf{\overline{X}}_{ji}^2=\sum_{i=1}^{M} \sum_{j=1}^{N_r}  \bar{n}_{ji}=N_r \left(\frac{M}{2} + \langle{O}\rangle\right).
\label{Mag_to_PCA}
\end{equation}
The above identity, together with Eq.~\eqref{SVD}, signifies that all the principal components reproduce exactly the expectation value of $\langle{O}\rangle$. As a consequence, 
we can rewrite Eq.~\eqref{Mag_to_PCA} as, 
\begin{equation}
    \langle{O}\rangle= \bar{\lambda}_1 + \Delta S - \frac{M}{2},
    \label{Universal_equation}
\end{equation}
where $\bar{\lambda}_1$ is the largest eigenvalue corresponding to $\mathbf{\overline{X}}$, the quantity $\Delta S = \sum_{i=2}^R \bar{\lambda}_i$ is the sum of all the eigenvalues, except $\bar{\lambda}_1$. 
Eq.~\eqref{Universal_equation} indicates that, for a given initial state, one can choose $a_i$, that maximizes the weight on $\bar{\lambda}_1$ compared to $\Delta S$ and, as a consequence, $\lambda_1$ alone could be enough to mimic the dynamics of the observable $O$, i.e., $\langle O(t) \rangle \approx \bar{\lambda}_1(t)$. In what follows, we show that for any choice of initial state, the best transformation that maximizes the weight on $\bar{\lambda}_1$ is when $a_i=\text{sgn}[\langle{S_i^z(0)}\rangle]$. We elucidate this point for the XXZ setup in Eq.~\eqref{XXZ-spin} for different operators and initial conditions, and show when the dominant eigenvalue $\bar{\lambda}_1$ provides the best estimate and why.

\textit{PCA with modified construction:-} We now present the results for the PCA performed on the modified snapshot matrix $\mathbf{\overline{X}}$ for different initial states of the XXZ chain. We choose the data set of size $N_r \times \frac{L}{2}$ and present here the results for $\Delta=1$.
As already seen from Fig.~\ref{fig:Bare_PCA}(b), for the DW initial state, $\lambda_1$ (which for the transformed case corresponds to $\bar{\lambda}_1$ for $a_i = +1$) contains most of the contribution relative to all the other $\lambda_k, (k>1)$, implying $\Delta S(t) \ll {\lambda}_1(t)$ in Eq.~\eqref{Universal_equation}, and thus accurately captures the dynamics of average magnetization $O = \sum_{i=1}^{L/2} S_i^z$ within the subsystem, as follows from Eq.~\eqref{Universal_equation} \cite{Devendra_PCA}. In fact, for the DW initial state, the other considered choices of $a_i \in \left[(-1)^i, [-1,1] \right]$, always lead to a smaller value of  $\bar{\lambda}_1(t)$ than that for $a_i=+1$ \cite{supp}. Here,  $[-1,1]$ implies that the values $\pm 1$ are chosen from a uniform distribution. Our result, therefore, indicates that $a_i= \mathrm {sgn} \big[\langle S_i^z(t=0)\rangle\big]$ gives the best choice for the PCA for the DW case. Moreover, for this choice of $a_i$, the connection of $\bar{\lambda}_1$ with the magnetization operator suggests a superdiffusive power-law growth of $\bar{\lambda}_1$ with a dynamical exponent $z=3/2$.
In Ref.~\cite{supp}, we further show results for the $\Delta \neq 1$ case, obtaining consistent results for spin transport.

\begin{figure}
    \centering
    \includegraphics[width=1\linewidth]{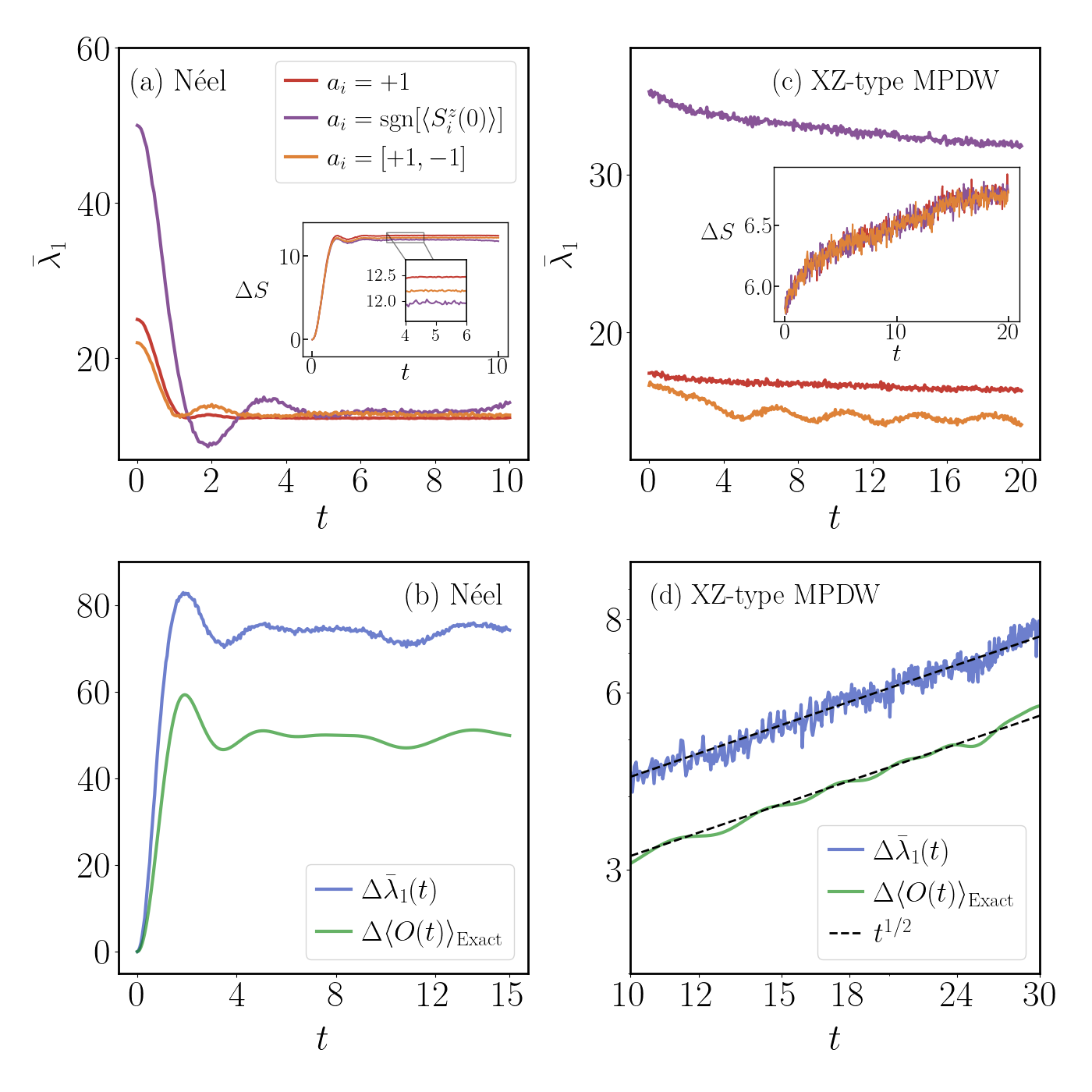}
    \caption{Plot of leading PCA eigenvalue $\bar{\lambda}_1$ for three different choices of $a_i \in \left[+1,\, \mathrm{sgn}[\langle S_i^z(0)\rangle], \, [-1,1]  \right]$, starting with the N\'eel state in (a\textbf{}) and XZ-type MPDW state in (c). The insets in (a) and (c) show the dynamics of $\Delta S(t)$, as defined in Eq.~\eqref{Universal_equation}, for the same three choices of $a_i$. In all cases, the choice $a_i = \mathrm{sgn}[\langle S_i^z(0) \rangle]$ leads to maximum weightage to  $\bar{\lambda}_1$.
    Plots (b), (d) shows growth of the exact $\Delta \langle{O(t)}\rangle_{\textrm{Exact}}=|\langle{O(t)}\rangle-\langle{O(0)}\rangle|$) (green solid line) and the leading PCA eigenvalue $\Delta \bar{\lambda}_1 =|\bar{\lambda}_1(t)-\bar{\lambda}_1(0)|$) (blue solid line) for the N\'eel and XZ-type MPDW initial states, respectively, where the form of $O$ is defined in Eq.~\eqref{adjusted_magnetization_defn} with $a_i = \mathrm{sgn}[\langle S_i^z(0)\rangle]$. $\Delta \bar{\lambda}_1$ captures the exact behaviour of the observable $O$ for both initial states and furthermore gives the correct dynamical exponent $z=2$ for the XZ-type MPDW initial states. The parameters chosen are the same as in Fig.~\ref{fig:Bare_PCA}.}
    \label{fig: Fig_2_Improved_PCA}
\end{figure}

We next consider the Néel state, and plot the dynamics of the principle component $\bar{\lambda}_1$ in Fig.~\ref{fig: Fig_2_Improved_PCA}(a), for different choices of $a_i \in \left[ +1, (-1)^i, [-1,1]  \right]$ which following Eq.~\eqref{adjusted_magnetization_defn} implies net magnetization, staggered magnetization, and a random linear combination of the local magnetization operators in a sub-system, respectively. We observe that for the case when $a_i=(-1)^i$, the weight on $\bar{\lambda}_1$ becomes the highest one. The inset shows that 
$\Delta S$ increases identically in all cases and finally saturates. However, the saturation is smallest for $a_i=(-1)^i$ which confirms that $a_i= \mathrm {sgn} \big[\langle S_i^z(t=0)\rangle\big]$ is the best estimator.  In other words, for the N\'eel state, the dynamics of $\bar{\lambda}_1$ can be well approximated by the exact dynamics of the staggered magnetization operator $O=\sum_{i=1}^{L/2} (-1)^i S_{i}^{z}$. Indeed, this is what we observe in Fig.~\ref{fig: Fig_2_Improved_PCA}(b), where we plot both the dynamics of $\bar{\lambda}_1$ as well as the exact evolution of the staggered magnetization operator. As seen, for the N\'eel state, the staggered magnetization increases rapidly and saturates, without providing a clear exponent that classifies the underlying transport. This implies that the transformed PCA on datasets starting from a Néel state captures only local correlations. In order to capture transport features starting with a N\'eel state, it is therefore important to look at higher order correlations \cite{fluctuating_hydrodynamics_in_chaotic_quantum_systems}. Given that the full snapshot matrix contains all order correlations, a natural question is how to construct a modified matrix that may allow the extraction of transport exponents using PCA. This will be discussed in the next section.

Finally, we consider the MPDW state.  Similar to the Néel state, the maximum weight on the PCA $\bar{\lambda}_1$ is once again obtained for $a_i=\text{sgn}\big[\langle S_{i}^{z}(0)\big \rangle]$, which in this case corresponds to the spin polarization \cite{universalscalinghigherordercumulants}. This is shown in Fig.~\ref{fig: Fig_2_Improved_PCA}(c), and we once again observe identical growth of $\Delta S$ for different choices of $a_i$, as shown in the inset.  Therefore, the exact dynamics of the spin-polarization operator is correctly captured via the transformed PCA.  We show this in Fig.~\ref{fig: Fig_2_Improved_PCA}(d). We observe a clear power-law growth in $\bar{\lambda}_1(t)$, which matches the exact growth of the spin polarization operator with dynamical exponent value $z=2$, indicating diffusive spin transport \cite{universalscalinghigherordercumulants,Non_eqb_dynamics_Spin_transport_in_a_tunable_Heisenberg_model_realized_with_ultracold_atoms}.

\textit{Higher-order correlations:-}  The PCA analysis performed on the data-sets (both the bare and modified) so far allows for capturing the local expectation values of operators of the form in Eq.~\eqref{adjusted_magnetization_defn}. However, it is not clear how the dynamics of non-local operators (higher-order correlations) can be extracted from the PCA analysis. In what follows, we provide two approaches to building the snapshot matrix, whose PCA analysis, in general, provides more information than just the expectation values of local observables.

One possible approach to extracting higher-order correlations is a natural generalization of the prescription presented above. For example, to calculate the second-order moment of the observable ${O}$ in Eq.~\eqref{adjusted_magnetization_defn}, i.e., $\langle{O^2}\rangle$, from the snapshot matrix $\mathbf{\overline{X}}$, one can proceed as follows. Given a snapshot vector $\boldsymbol{\bar{n}}=(\bar{n}_1, \bar{n}_2,..., \bar{n}_M)$ in Eq.~\eqref{Transformed_snap_new_observable}, we define the second order snapshot vector as $(\bar{n}_i + \bar{n}_j)$ mod $2$, which has a dimension of $M^2$. Recall that $M$ is the domain size. All these second-order snapshots are arranged as row vectors to form a rectangular matrix $\overline{\mathbf{Z}}$ of dimension $N_r \times M^2$. As the entries of the matrix $\overline{\mathbf{Z}}$ are $0$ or $1$, the same fundamental theorem of the SVD relates the $\langle{O^2}\rangle$ to the singular values of the matrix $\overline{\mathbf{Z}}$ as \cite{supp}
\begin{equation}
\langle{O^2}\rangle = \frac{M^2}{4} - \frac{\Lambda_1}{2} -  \frac{\sum_{k=2}^R \Lambda_k} {2}.
\label{second-order}
\end{equation}
Therefore, using the first principal component of $\overline{\mathbf{X}}$ and $\overline{\mathbf{Z}}$ one can express the second-order cumulant as, 
\begin{equation}
\sigma^2_{\textrm{PCA}}(t) = M\bar{\lambda}_1(t) -\bar{\lambda}^2_1(t) -\frac{\Lambda_1(t)}{2} + \mathrm{other \, terms},
\label{PCA_second_order_cumulant}
\end{equation}
where $\Lambda_1 (\bar{\lambda}_1)$ is the principal component for the first (second) order snapshot matrix. While the sum of all PCA eigenvalues exactly reproduces $\langle{O^2}\rangle$ [Eq.~\eqref{second-order}], the first principal component $\Lambda_1$ of the matrix $\overline{\mathbf{Z}}$ can contain the majority of the information where the other terms (that depends on all the other eigenvalues) can be neglected and therefore could be enough to mimic the dynamics of higher-order correlations \cite{supp}. In the following, we restrict ourselves to the first component, $\Lambda_1$, and examine whether it is sufficient to capture the dynamics of nonlocal correlations. Note that this approach can be systematically extended to study  $\langle{O^n}\rangle$, $n>2$ \cite{supp}.

\begin{figure}
    \centering
\includegraphics[width=1\linewidth]{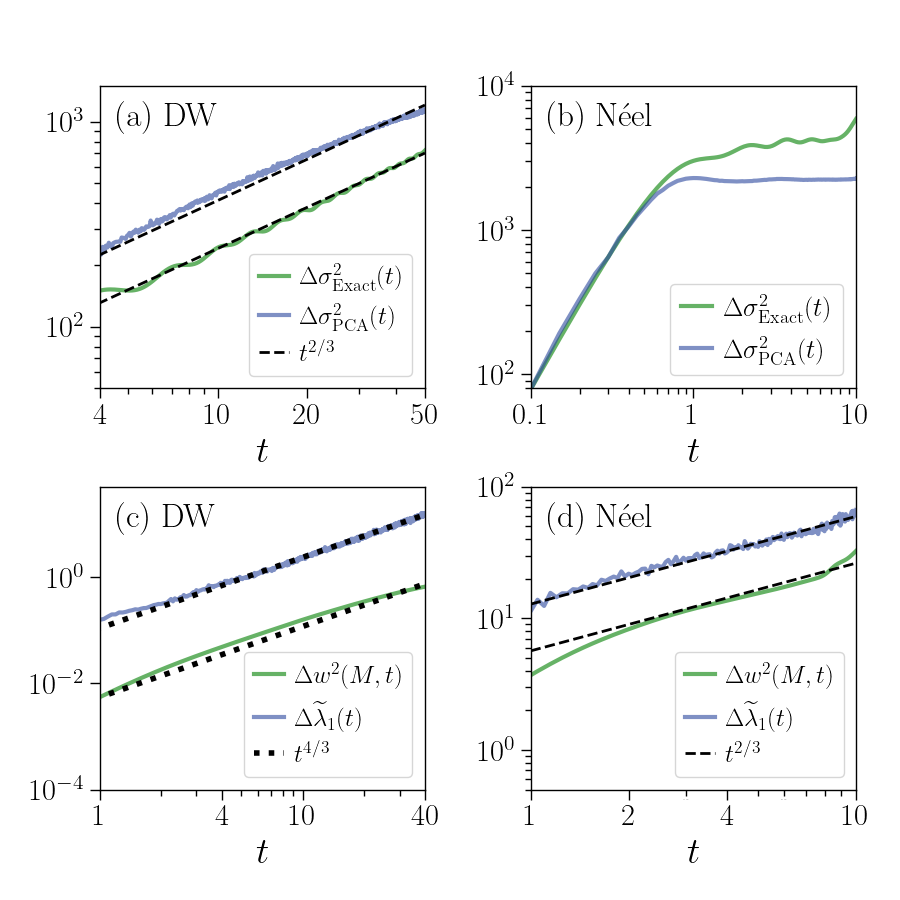}
    \caption{Plot for second order cumulant $\sigma_{\textrm{Exact}}^2(t)=\langle O^2(t)\rangle-\langle O(t) \rangle^2$, with $a_i=\mathrm{sgn}[\langle S_i^z(0)\rangle]$. Both PCA approximated dynamics $\Delta \sigma^2_{\textrm{PCA}}(t)=|\sigma^2_{\textrm{PCA}}(t)-\sigma^2_{\textrm{PCA}}(0)|$ [Eq.~\eqref{PCA_second_order_cumulant}] and the exact dynamics $\Delta \sigma^2_{\textrm{Exact}}(t)=|\sigma^2_{\textrm{Exact}}(t)-\sigma^2_{\textrm{Exact}}(0)|$  are shown for DW initial state in (a) and for N\'eel state in (b). The PCA approach captures the $z=3/2$ dynamical exponent in the DW case, whereas no exact behaviour can be extracted from the N\'eel initial state. Exact growth of $\Delta w^2(M,t)= |w^2(M,t) - w^2(M,0)|$, and the first principal component $\Delta \widetilde{\lambda}_1(t)=| \widetilde{\lambda}_1(t)-\widetilde{\lambda}_1(0)|$ obtained from the snapshot matrix defined in Eq.~\eqref{Snapshot_matrix_Surface_Roughness} are plotted for the DW initial state in (c) and the N\'eel state in (d), respectively. The first principal component $\widetilde{\lambda}_1(t)$ captures the same dynamical exponent as the exact $w^2(M,t)$, given by $z=3/4$ for the DW case, and $z=3/2$ for the Néel case. The parameters used here are $L=100$, $N_r=1000$, and the domain $M$ is chosen from the 10th to the 90th site of the lattice chain. In all plots, the quantity $\Delta\sigma^2_{\textrm{Exact}}(t)$ and $\Delta w^2(M,t)$ is multiplied by a constant factor $d$ for plotting it on the same scale as $\Delta\sigma^2_{\textrm{PCA}}(t)$ and $\Delta \widetilde{{\lambda}}_1(t)$ respectively. The value of $d$ is $100$ for (a) and (b), $0.1$ for (c), and $40$ for (d).}
    \label{fig:Higher_order}
\end{figure}

We now present the numerical results to capture the dynamics of $\sigma^2_{\textrm{Exact}}(t) = \langle O^2 (t) \rangle - \langle O (t) \rangle^2$ following the PCA analysis that is performed on datasets $\mathbf{\overline{X}}(t)$ and $\overline{\mathbf{Z}}(t)$.  We first consider the DW case and in Fig.~\ref{fig:Higher_order}(a), we plot the exact dynamics of $\sigma^2_{\textrm{Exact}}(t)$, which corresponds to growth of magnetization fluctuation in a sub-system, for the XXZ chain for $\Delta=1$ and compare it with the one obtained following the PCA, where we plot $\sigma^2_{\textrm{PCA}}(t)$ by approximating it by the first three terms in Eq.~\eqref{PCA_second_order_cumulant}. We find that both the exact and the PCA-approximated dynamics grow as a power law with dynamical exponent $z=3/2$, highlighting the validity of our PCA analysis in capturing the correct higher-order correlations for the DW initial state. It is important to note that for this case, ``other terms'' in Eq.~\eqref{PCA_second_order_cumulant} can be neglected \cite{supp}.

We next consider the dynamics starting from the Néel state, and the result is shown in Fig.~\ref{fig:Higher_order}(b). We find that the higher-order construction, presented above, does not capture the exact dynamics and the underlying transport features.  This, however, is expected, as the exact dynamics here correspond to fluctuations in the staggered magnetization $[a_i=(-1)^i$ in Eq.~\eqref{adjusted_magnetization_defn}], which do not carry transport signatures. Furthermore, the reason for the mismatch between PCA and exact is due to the similar order of magnitude of $\Lambda_1$ and $\Delta S$, leading to a non-negligible contribution of the ``other terms'' in Eq.~\eqref{PCA_second_order_cumulant}. For further details, please see Ref.~\cite{supp}.

We now propose a new construction that enables PCA to capture transport features, even for a N\'eel initial state. The modified data set is constructed by replacing each element with the cumulative sum of the elements in the same row up to that column. Formally,
\begin{equation}
\widetilde{X}_{i\ell}(t) = \sum_{k=1}^{\ell} n_i^{k},
\quad i = 1,\dots, N_r, \quad \ell = 1,\dots,M ,
\label{Snapshot_matrix_Surface_Roughness}
\end{equation}
where $n_{i}^{k}$ denotes the element of the original data set $\mathbf{X}(t)$ (bare dataset) corresponding to the $i$-th row and the $k$-th column.
Importantly, with this construction, we can relate the sum of the eigenvalues of the data matrix to quantum surface roughness, which is known to exhibit power-law behavior \cite{FV_Scaling_of_Roughness_Growth_Strongly_Interacting_Bose_Gas,Dynamical_Scaling_Surface_Roughness_and_Entanglement_Entropy_Disordered_Fermion_Models,fermion_bipartite_fluctuations,FV_Scaling_and_KPZ_growth_surface_roughness_in_driven_one_dimensional_quasiperiodic_model,Surface_roughness_Impact_of_Dissipation_on_Universal_Fluctuation_Dynamics_in_Open_Quantum_Systems}. For the quantum surface roughness, we first define a quantum surface height operator as ${h}_{j}= \sum_{i=1}^{j} S_z^{i}$. This operator is a sum of local particle-number operators at the same time. For a given length scale (height), this operator gives the magnetization fluctuations summed up to the site $j$, relative to a reference level. The average surface height is
given by $h_{\text{av}}=\frac{1}{M}\sum_{j=1}^{M}\text{Tr}[{\rho}(t) {h}_{j}].$
The square of surface roughness is then defined as the expectation value of the variance of the quantum surface height operator as
\begin{align}
w^2(M,t)= & {\frac{1}{M}\sum_{j=1}^{M}\text{Tr}\left[{\rho}(t)[{h}_{j}-h_{\text{av}}(t)]^{2}\right]}.
\label{Exact_Surface_roughness}
\end{align}
In Fig.~\ref{fig:Higher_order}(c,d), we perform the PCA on the modified data-set $\widetilde{X}(t)$ and plot the dynamics of $\widetilde{\lambda}_1(t)$ for both DW and N\'eel state together with the exact dynamics for the surface roughness $w^2(M,t)$, given in Eq.~\eqref{Exact_Surface_roughness}. With this construction, the dynamics of $\widetilde{\lambda}_1(t)$ starting from the N\'eel state now grow as $\sim t^{2/3}$ and therefore, capture the information about the underlying transport. Interestingly, for the DW case, we observe a faster growth of $\widetilde{\lambda}_1(t) \sim t^{4/3}$. Such different power laws are, in fact, the feature of the exact dynamics of the surface roughness, and this gets captured by the PCA 
$\widetilde{\lambda}_1(t)$, suggesting that it captures the information of the surface roughness observable.

\textit{Summary.--}
We investigate non-equilibrium quantum dynamics using PCA and applying the method to wavefunction snapshot data. We show that the information content of PCA depends sensitively on the choice of initial state and the data representation. While the full dynamics may be distributed across all components, an appropriate transformation of the snapshot matrix can concentrate the maximal information into the largest principal component. This transformation directly identifies the observable whose dynamics are encoded in the principal component, allowing the accurate extraction of the dynamical exponents of the underlying transport. 
We further extend our analysis to capture non-local correlations via PCA by providing an appropriate transformation of the snapshot matrices. We present all the findings for a one-dimensional XXZ Heisenberg spin chain, starting from different initial states. 

Our work demonstrates the applicability of PCA to study various aspects of non-equilibrium dynamics, from thermalization to quantum transport, and clarifies which operator is closely related to the PCA outcomes, offering a clear physical interpretation of dimensionality-reduction techniques.

\textit{Acknowledgement.--} DSB
acknowledges Marcello Dalmonte and R. Verdel for useful discussions and for collaborations on related projects. 
DY and BKA acknowledge the National Supercomputing Mission (NSM) for providing computing resources of ‘PARAM Brahma’ at IISER Pune, which is implemented by C-DAC and supported by the Ministry of Electronics and Information Technology (MeitY) and DST, Government of India. BKA acknowledges CRG Grant No. CRG/2023/003377 from Science and Engineering Research Board (SERB), Government of India.  BKA acknowledges the hospitality of The Abdus Salam International Center for Theoretical Physics (ICTP), Italy, under the Associateship Program.

\bibliography{main_references.bib}

\setcounter{equation}{0}
\setcounter{figure}{0}
\renewcommand{\theequation}{S\arabic{equation}}
\renewcommand{\thefigure}{S\arabic{figure}}

\onecolumngrid
\newpage


\begin{center}
{\textbf{{Supplemental Material: Principal component analysis of wavefunction snapshots in non-equilibrium dynamics}}}
\end{center}


\section{Details about the Multi-Periodic Domain Wall (MPDW) initial state}
\label{initialstate}
In this section, we provide details on the multi-periodic domain wall (MPDW) state \cite{universalscalinghigherordercumulants}, which we have considered as one of the initial states for studying exact dynamics and PCA-approximated dynamics. MPDW state is an extension of the domain wall (DW) state, where each local qubit exists in a superposition state to create a global product state. The DW initial state consists of a half-chain spin-up and a half-chain spin-down configuration 
\begin{equation}
    \ket{\psi_{DW}} = \ket{\uparrow} ^{\otimes L/2} \otimes \ket{\downarrow} ^{\otimes L/2}.
\label{Domain_wall_form}
\end{equation}
The general local wavefunction at site $i$ of a multi-periodic domain-wall (MPDW) state has the form,
\begin{equation}
\ket{\psi_{\mathrm{MPDW}}}_i =
\begin{cases}
A \ket{\uparrow}_i + B \ket{\downarrow}_i,
& \cos(Q i + \theta) > 0 \ \text{and} \ \cos(Q i + \theta + \phi) > 0, \\
A \ket{\uparrow}_i - B \ket{\downarrow}_i,
& \cos(Q i + \theta) > 0 \ \text{and} \ \cos(Q i + \theta + \phi) < 0, \\
A \ket{\downarrow}_i + B \ket{\uparrow}_i,
& \cos(Q i + \theta) < 0 \ \text{and} \ \cos(Q i + \theta + \phi) > 0, \\
A \ket{\downarrow}_i - B \ket{\uparrow}_i,
& \cos(Q i + \theta) < 0 \ \text{and} \ \cos(Q i + \theta + \phi) < 0,
\end{cases}
\label{Genreal_MPDW_state}
\end{equation}
where, $Q$ is the wave-vector, defined as $Q=2\pi/\lambda$, $\theta$, and $\phi$ are the phases, $A = \sqrt{\frac{k+1}{2k}}$, $B = \sqrt{\frac{k-1}{2k}}$, with  $k = \sqrt{\eta^2 + 1}$, $\eta \in[0,\infty)$. The global initial state is given by the direct product of the local wavefunctions $\ket{\psi_{MPDW}}= \otimes_i\ket{\psi_{{MPDW}}}_i$. The above MPDW state mimics a DW state when $\eta=0$, which suppresses the $x$ component, resulting in a configuration that has all spins pointing in the $z$ direction.

The XZ-type MPDW state is a special case of the general MPDW state, which has the superposition along both $x$ and $z$ bases compared to just $z$ basis in the DW. It has parameters $\eta =1$ and $\phi = \pi/2$, such that it has equal amplitudes in both $x$ and $z$ components of the local spins, 

\begin{equation}
\ket{\psi_{XZ}}_i =
\begin{cases}
\sqrt{\dfrac{\sqrt{2}+1}{2\sqrt{2}}}\ket{\uparrow}_i
+ \sqrt{\dfrac{\sqrt{2}-1}{2\sqrt{2}}}\ket{\downarrow}_i,
& \cos(Q i + \theta) > 0 \ \text{and} \ \cos(Q i + \theta + \dfrac{\pi}{2}) > 0, \\[6pt]
\sqrt{\dfrac{\sqrt{2}+1}{2\sqrt{2}}}\ket{\uparrow}_i
- \sqrt{\dfrac{\sqrt{2}-1}{2\sqrt{2}}}\ket{\downarrow}_i,
& \cos(Q i + \theta) > 0 \ \text{and} \ \cos(Q i + \theta + \dfrac{\pi}{2}) < 0, \\[6pt]
\sqrt{\dfrac{\sqrt{2}+1}{2\sqrt{2}}}\ket{\downarrow}_i
+ \sqrt{\dfrac{\sqrt{2}-1}{2\sqrt{2}}}\ket{\uparrow}_i,
& \cos(Q i + \theta) < 0 \ \text{and} \ \cos(Q i + \theta + \dfrac{\pi}{2}) > 0, \\[6pt]
\sqrt{\dfrac{\sqrt{2}+1}{2\sqrt{2}}}\ket{\downarrow}_i
- \sqrt{\dfrac{\sqrt{2}-1}{2\sqrt{2}}}\ket{\uparrow}_i,
& \cos(Q i + \theta) < 0 \ \text{and} \ \cos(Q i + \theta + \dfrac{\pi}{2}) < 0 .
\end{cases}
\label{XZ_type_MPDW_state}
\end{equation}

Therefore, the global XZ-type MPDW initial state is, $\ket{\psi_{XZ}}= \otimes_i \ket{\psi_{XZ}}_i$, with a wavevector $Q=2\pi/\lambda_{XZ}$.

\section{Saturation of information content in the dominant PCA component}
\label{sec: Saturation_of_information_content}
In this section, we examine the saturation value of the first principal component $\lambda_1$, once thermalization is achieved for a fully chaotic system. For simplicity, we consider an infinite-temperature state with the form, 
\begin{equation}
    \ket{\psi}_\text{thermal} = \frac{1}{\sqrt{D}}\sum_{k=1}^{D} \ket{k}
    \label{Infinte_temp_state}
\end{equation}
where $D=2^L$ is the dimension of the Hilbert space with $L$ being the system size. One can think of all the $2^L$ possible configurations to lie at the vertices of an $L$-dimensional hypercube. For the given state in Eq.~\eqref{Infinte_temp_state}, each configuration is equally likely to occur in measurement. Therefore, we  construct the snapshot matrix $\mathbf{X}$ consisting of a total $N_r=n \times 2^L$ realizations, where $n$ corresponds to number of occurrence of each of the $2^L$ possible configurations. 

We now proceed further and calculate the total variance of the snapshot matrix $\mathbf{X}$, which we denote as $V(\mathbf{X})$. If $P_{ji}$ is the projection of the $j$-th snapshot $(j=1, 2, \cdots N_r)$ on the $i$-th axis $(i=1, 2, 3, \cdots L)$ of the $L$ dimensional hypercube, then $V(\mathbf{X})$ is given as,
\begin{eqnarray}
V(\mathbf{X}) &=& \frac{1}{N_r} \sum_{i=1}^{L} \sum_{j=1}^{N_r} P_{ji}^2 \nonumber \\
&=& \frac{n  \sum_{l=0}^L  \, l \,  \binom{L}{l}}{n \times 2^L} = \frac{L \, 2^{L-1}\cdot}{2^L} = \frac{L}{2},
\label{variance_of_X}
\end{eqnarray}
where in the second line, we count number of realizations for which fixed $l$ number of times the value 1 occurs and remaining $L-l$ number of times 0 occurs. That gives the combinatorial factor in the numerator and such occurrence will happen $n$ times. 

Let us now consider a line along the diagonal of the hypercube, joining the points $(0,0,\cdots,0)$ and $(1,1,\cdots,1)$ which we call as one of the principal axes. The principal eigenvalue $\lambda_p$ along this principal direction is $ \frac{1}{N_r} \sum_{j=1}^{N_r} P_j^2 $, where $P_j$ is the projection of the $j$-th state onto the diagonal axis. This is given as 
\begin{equation}
    \lambda_p = \frac{\sum_{j=1}^{N_r} P_j^2}{n \times 2^L} = \frac{n \, \sum_{l=0}^L  l^2 \binom{L}{l}} {n \times L \times 2^L} = \frac{L(L+1)2^{L-2}}{L\times 2^L} = \frac{L+1}{4}. \label{lambda_1_most_delocalized_state}
\end{equation}
Using \eqref{variance_of_X} and \eqref{lambda_1_most_delocalized_state}, we can find the amount of variance contained in the principal component relative to the total variance is 
\begin{equation}
    \frac{\lambda_p}{V(\mathbf{x})} = \frac{(L+1)/4}{L/2} = \frac{1}{2} \Big[1+ \frac{1}{L}\Big].
    \label{lambda_1_ratio_random_state}
\end{equation}
Therefore, we find that there exist a direction that captures more than 50\% of the total variance, making it the first principal component, i.e., $\lambda_p=\lambda_1$. As $L \to \infty$, the first principal component captures exactly 50\% of the total variance.

The above calculation only sets a system-size limit for the ratio $\frac{\lambda_1}{V(\mathbf{X})}$ in a random infinite-temperature state for a fully chaotic system. In the presence of any additional symmetries or conservations in the Hilbert space, the ratio can saturate at any value between 0 and 1, depending on the value of the conserved quantity.

\section{Dynamics of ${\Delta S}$ in Eq.~(\ref{Universal_equation})}
\label{Supp_Sec: Universality_of_Delta_S}

\begin{figure}[h!]
    \centering
    \includegraphics[width=0.7\textwidth]{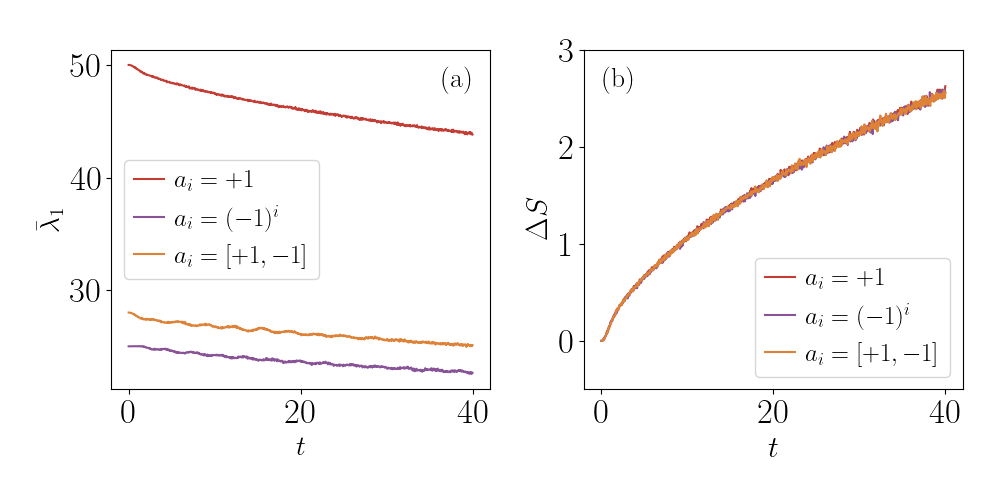}
    \caption{Plot for the dynamics of (a) $\bar{\lambda}_1$, and (b) $\Delta S$ for the Domain wall (DW) initial state. We observe that $\bar{\lambda}_1$ corresponding to $a_i=+1$ has the highest weightage compared to other choices of $a_i$. The $\Delta S$, however, in all cases grows identically. This confirms that $\bar{\lambda}_1$ best approximates the dynamics of observable $\langle O \rangle$.  For the numerics we consider, $\Delta=1$, $L=100$, and $N_r=1000$.}
    \label{fig: Supp_Delta_E}
\end{figure}

In this section, we study the dynamics of $\Delta S$, defined in Eq.~\eqref{Universal_equation}, and provide arguments in support of the
the best transformation of the snapshot matrix, as proposed in the main text.  Given a general operator ${O} = \sum_{i=1}^M a_i S_i^z$, we express its expectation value as  $\langle{O(t)}\rangle=\bar{\lambda}_1(t)+\Delta S(t) - M/2$ [Eq.~\eqref{Universal_equation}], where recall that  $\bar{\lambda}_1$ in the principal component of the transformed snapshot matrix $\overline{\mathbf{X}}$ and $\Delta S= \sum_{i=2}^{R} \bar{\lambda}_i$. Here we show that, for the Heisenberg spin chain, starting with  the DW initial condition, the choice $a_i=+1$ maximizes the weight on $\bar{\lambda}_1$. To demonstrate this, we consider three different choices of $a_i$ which are $a_i=+1$, $a_i=(-1)^i$, and $a_i$ uniformly chosen from $[+1,-1]$, and the plot the dynamics of   $\bar{\lambda}_1$ and $\Delta S$ and in Fig.~\ref{fig: Supp_Delta_E}, following the transformed snapshot matrix $\overline{\mathbf{X}}$. We see in Fig.~\ref{fig: Supp_Delta_E} (a) that $a_i=+1$ have the highest weightage for $\bar{\lambda}_1$ compared to the other two choices of $a_i$. From Fig.~\ref{fig: Supp_Delta_E}(b), it can be clearly inferred that all three curves of $\Delta S$ corresponding to the three choices of $a_i$ evolves identically.


The identical growth of $\Delta S$ suggests that the $\bar{\lambda}_1$ best approximates $\langle O \rangle$, when the $\langle O \rangle \gg\Delta S$. It can be intuitively inferred from the following numerical argument. The snapshot matrix can have only two possible extreme values: all entries are either $0$ or $1$. In the physical spins, by multiplying the spins with their opposite sign, it reproduces the values $+1/2$, making the snapshot entries closer to 1 (in our mapping), thereby increasing the expectation value of the observable $O$. As $\Delta S$ is constant across these different operators $O$, the principal component $\bar{\lambda}_1$ should increase to maintain the equality in Eq.~\ref{Universal_equation}. This can be seen in the plots of Fig. \ref{fig: Fig_2_Improved_PCA} (a) and (b), which clearly show that the $\bar{\lambda}_1$ corresponding to the spin polarization captures the maximum information. This behavior can also be seen in the Fig.~\ref{fig: Supp_Delta_E} (a), where $a_i=+1$ have the highest value, as $\mathrm{sgn}[S_i^z(0)]=+1$, for $i=1,2, \cdots, L/2$.

Now, for capturing the dynamics of non-local operators, the static and dynamic nature of $\Delta S$ and $\bar{\lambda}_1$ determine the number of terms required to approximate the second-order cumulant (or variance). Expressing the variance $\sigma^2 (t)$ in terms of the principal components gives us the form of `other terms' in Eq. ~\eqref{PCA_second_order_cumulant}. We write, 
\begin{equation}
    \sigma ^2_{\textrm{PCA}} (t) = \left[M\bar{\lambda}_1(t)-\frac{\Lambda_1(t)}{2} - \bar{\lambda}_1^2(t) \right] +  \left[M\Delta S_{(1)}(t)-\frac{\Delta S_{(2)}(t)}{2} - \Delta S^2_{(1)}(t)  - 2\bar{\lambda}_1(t)\Delta S_{(1)}(t) \right],
\end{equation}
where $\Delta S_{(1,2)}$ are the sum of the rest of the principal eigenvalues for the first order $\mathbf{\overline{X}}$ and second order $\overline{\mathbf{Z}}$ snapshot matrix. When $\Delta S_{(1,2)}$ is very small compared to $\bar{\lambda}_1$ and $\Lambda_1$, we can neglect the other terms, as in the case of DW. However, each of these terms becomes important when the scale of $\Delta S_{(1,2)}$ is comparable to $\bar{\lambda}_1$ and $\Lambda_1$, which is the case, for example, for the N\'eel state.

\section{Applicability of PCA  for XXZ chain for $\Delta <1$ and $\Delta>1$ regime}
\begin{figure*}[t]
    \centering    \includegraphics[width=0.95\textwidth]{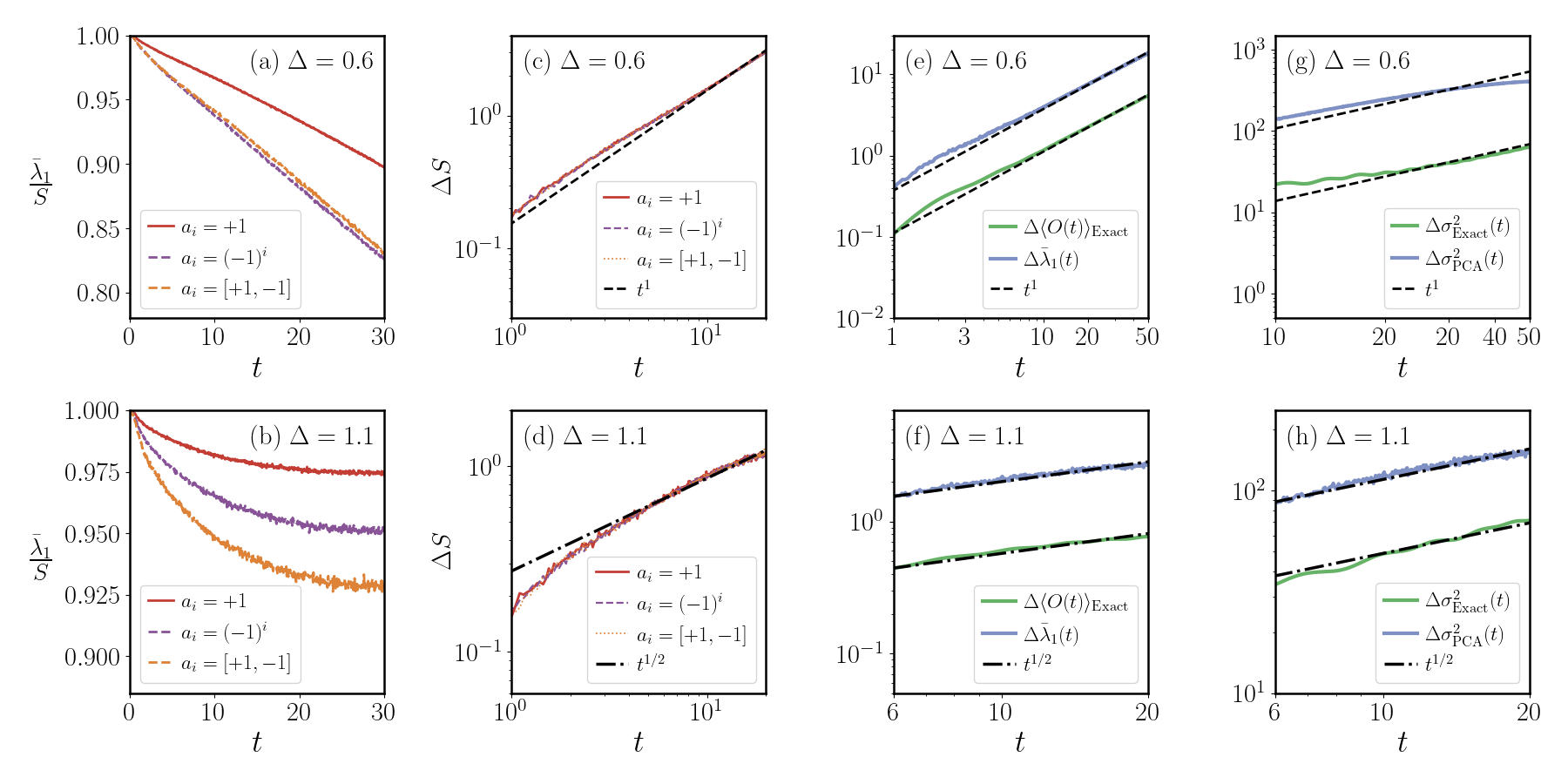}
    \caption{(a)-(b) Plot for scaled eigenvalue $\bar{\lambda}_1$ (scaled by $S=\sum_{i} \bar{\lambda}_i$) for three different choices of $a_i$ for the XXZ spin chain 
    for $\Delta = 0.6$ (easy-plane) in (a) and $\Delta =1.1$ (easy-axis) in (b), starting from the DW initial state.  (c)-(d) Plot for $\Delta S$ which evolves almost identically for all the three choices of $a_i$ for both the $\Delta$ values. (e)-(f) Exact dynamics of $\Delta \langle O(t) \rangle_{\textrm{Exact}} =|\langle O(t) \rangle -\langle O(0) \rangle|$ (scaled by a scalar $d$) obtained via TEBD and first principal component $\Delta \bar{\lambda}_1(t) = |\bar{\lambda}_1(t) - \bar{\lambda}_1(0)|$ of the snapshot matrix $\overline{\mathbf{{X}}}$. The correct transport exponents are captured in $\Delta \lambda_1(t)$ as in the $\Delta \langle O(t) \rangle_{\textrm{Exact}}$, which are ballistic for  $\Delta=0.6$ as shown in (e), whereas diffusive for $\Delta=1.1$ as shown in (f). In (g) and (h), the dynamics of the second-order cumulant of observable $O$ defined as $\sigma^2_{\textrm{Exact}}(t) = \langle{O^2(t)}\rangle - \langle{O(t)}\rangle^2$, when $a_i=+1$ is shown. Both PCA $\Delta \sigma^2_{\textrm{PCA}}(t)=|\sigma^2_{\textrm{PCA}}(t)-\sigma^2_{\textrm{PCA}}(0)|$ and exact $\Delta \sigma^2_{\textrm{Exact}}(t)=|\sigma^2_{\textrm{Exact}}(t)-\sigma^2_{\textrm{Exact}}(0)|$ (scaled by $d$) are shown in (g) and (h), where $\sigma^2_{\textrm{PCA}}(t)$ is defined in Eq.~\eqref{PCA_second_order_cumulant}. $\bar{\Lambda}_1$ corresponding to $\mathbf{\bar{Z}}$ captures the ballistic scaling in (g) and the diffusive scaling in (h). For the numerics, we use $L=100$, and the number of realizations $N_r=1000$. The scaling factor $d$ is $0.5$ for (e) and (f), $50$ for (g), and $80$ for (h).}
    \label{fig: Ansitropic_spin_chain}
\end{figure*}

In this main text, we have provided results for the isotropic XXZ chain $(\Delta =1)$. In this section, we demonstrate the validity of the PCA technique for the anisotropic case $(\Delta \neq 1)$. We primarily focus on the domain-wall (DW) initial state. In the main text, in the isotropic case ($\Delta=1$), we observed a clear super-diffusive exponent in the first principal component [see Fig.~\ref{fig:Bare_PCA}]. A similar PCA analysis is performed here for the easy-plane ($\Delta < 1$) and the easy-axis ($\Delta > 1$) regime, which exhibit ballistic and diffusive transport, respectively, for the principal component. We see that the largest amount of the information $\bar{\lambda}_1$, corresponding to the snapshot matrix $\mathbf{\overline{X}}$ is captured when $a_i=+1$, compared to the other two cases i.e., $a_i=(-1)^i$ or $a_i=[+1,-1]$ for both the easy-plane and the easy-axis regimes. As a result, the operator $O$ in Eq.~\eqref{adjusted_magnetization_defn}, is best estimated by $\bar{\lambda}_1$ for $a_i=1$ which corresponds to magnetization operator $O= \sum_{i=1}^{L/2} S_z^i$ of the subsystem of size $L/2$.  We show this in Fig.~\ref{fig: Ansitropic_spin_chain}(a) and (b) for  $\Delta=0.6$, and  $\Delta=1.1$, respectively, where we plot scaled $\bar{\lambda}_1$ (scaled by the quantity $S=\sum_{i} \bar{\lambda}_i$) for three different choices of $a_i$. We observe that $\bar{\lambda}_1$ corresponding to $a_i=+1$ has the highest weightage. It can also be seen in Fig.~\ref{fig: Ansitropic_spin_chain}(c) and (d) that the $\Delta S$ evolves almost identically for all three choices of $a_i$ for both $\Delta < 1$ and $\Delta > 1$ regimes, respectively. 
This behavior is in accordance with the discussion in sec \ref{Supp_Sec: Universality_of_Delta_S}. 

Fig.~\ref{fig: Ansitropic_spin_chain}(e) and (f) show the first-order expectation value of magnetization in the left half chain, and the $\bar{\lambda}_1$ perfectly captures the ballistic and diffusive exponents, respectively, and matches the predictions that follow from the exact dynamics.  Fig.~\ref{fig: Ansitropic_spin_chain} (g) and (h) show the second-order cumulant of $O$, which corresponds to particle number fluctuations in the left half of the chain. Ballistic growth is observed for the easy-plane ($\Delta =0.6$), whereas easy-axis ($\Delta =1.1$) shows diffusive $z=2$  growth for both magnetization and particle number fluctuations. We observe that ${\Lambda}_1$ computed following the snapshot matrix $\mathbf{\overline{Z}}$, as discussed in the main text, matches the prediction that follows from the exact dynamics. 

\section{Snapshot matrix construction for higher-order correlations}
In this section, we provide details on constructing the snapshot matrix to calculate higher-order (non-local) correlations. We consider the generic operator $O = \frac{1}{2}\sum_{i=1}^{M} a_i\sigma_z^i$, where $a_i$ is a real scalar factor having values either $-1$ or $+1$. In the main text, we have shown that the snapshot matrix $\mathbf{{X}}$ can be transformed to form a new matrix $\overline{\mathbf{{X}}}$ [Eq.~\eqref{Transformed_snap_new_observable}] whose classical average reproduces the expectation value of $\langle O \rangle$ [ Eq.~\eqref{Mag_to_PCA}]. Recall that the transformation being 
$\bar{n}_i = n_i \oplus \bar{a}_i$, where $\bar{n}_i$ are the transformed snapshots and the mapping between physical spin and computational symbol for all the snapshots considered in this study is $\sigma_z^i\ket{0}=-1\ket{0}, \sigma_z^i\ket{1}=+1\ket{1}$.  
With this convention, starting from the local operator $O$, the second order operator will have the form $O^2 = \frac{1}{4} \sum_{i,j=1}^{M} a_i a_j \sigma_z^i \sigma_z^j$, which is non-local in nature, and whose expectation value can be written as,
\begin{equation}
    \langle{\psi(t)|O^2|\psi(t)}\rangle = \sum_{k=1}^{D} \sum_{i,j=1}^{M} \frac{|c_k|^2}{4}\langle{k|a_ia_j\sigma_z^i\sigma_z^j|k}\rangle,
\end{equation}
where the state $\ket{\psi(t)} = \sum_{k=1}^{D} c_k\ket{k}$, with $D$ being the Hilbert space dimension. The basis state $\ket{k}$ is a binary string of length $M$, which creates a row of the snapshot matrix. The operator $a_i\sigma_z^i$ acting on $\ket{k}$ reproduces the snapshot matrix element $\bar{n}_i$, using which we can construct the second-order snapshot matrix $\overline{\boldsymbol{Z}}$, corresponding to $O^2$, in the following way,
\begin{equation}
    \bar{Z}_{ij} = (\bar{n}_i + \bar{n}_j) \bmod 2, \quad \quad i,j=1, 2, \cdots, M
    \label{barZ}
\end{equation}    
where $\bar{n}_i$ and $\bar{n}_j$ are the transformed snapshots at sites $i$ and $j$ respectively. By repeating this procedure over $N_r$ realizations and arranging each $\bar{Z}$ in Eq.~\eqref{barZ} in one row, the second order snapshot matrix $\overline{\boldsymbol{Z}}$ is obtained, which satisfies the following identity 
\begin{equation}
    \sum_{k=1}^D \sum_{i,j=1}^{M} \frac{|c_k|^2}{4}\langle{k|a_ia_j\sigma_z^i\sigma_z^j|k}\rangle = \sum_{m=1}^{N_r}\frac{1}{2 \cdot N_r}\left[\frac{M^2}{2} - \sum_{l=1}^{M^2}\overline{\boldsymbol{Z}}_{ml}\right],
\end{equation}
which further implies 
\begin{equation}
   \langle{\psi(t)|O^2|\psi(t)}\rangle =  \frac{1}{2}\left[\frac{M^2}{2} - \frac{\sum_{m=1}^{N_r} \sum_{l=1}^{M^2} \overline{\boldsymbol{Z}}_{ml}}{N_r}\right]. 
\end{equation}
Employing the fundamental theorem of the single value decomposition (SVD), given as, $\sum_{k=1}^Rs_k^2(\bar{Z}) = \sum_{m=1}^{N_r}\sum_{l=1}^{M^2} \bar{Z}_{ml}$,  we receive the relation between all the PCA components and the exact value of $\langle O^2(t) \rangle$ as
\begin{equation}
    \langle{\psi(t)|O^2|\psi(t)}\rangle =  \frac{M^2}{4} - \frac{\sum_{k=1}^R\Lambda_k(\overline{\boldsymbol{Z}})}{2},
    \label{exact_pca_final_relation}
\end{equation}
which is given in Eq.~\eqref{second-order} of the main text. Here recall that $R$ is the rank of the matrix $\mathbf{\bar{Z}}$. Interestingly, the above prescription can be systematically extended to compute expectation values of higher orders of operators,  
\begin{equation}
    \langle{\psi(t)|O^n|\psi(t)}\rangle = \frac{(-1)^n}{2^{n-1}}\left[\frac{M^n}{2} - \sum_{k=1}^{R} {\Lambda_k(\overline{\boldsymbol{Z}}_{N_r \times M^n})}\right],
\label{exact_pca_higher_order_relation}
\end{equation}
where $\overline{\boldsymbol{Z}}_{N_r \times M^n}$ is the snapshot matrix constructed by extending the above described scheme.

\end{document}